\documentclass[aps,pra,twocolumn,notitlepage,10pt]{revtex4-2}
\usepackage[utf8]{inputenc}
\usepackage{float}
\usepackage{blindtext}
\usepackage{amsmath}
\usepackage{amssymb}
\usepackage{graphicx}
\usepackage{xfrac}

\usepackage[dvipsnames]{xcolor}
\usepackage{epstopdf}

\usepackage{dsfont}
\usepackage{enumitem}
 \usepackage{hyperref} 
\usepackage[capitalise]{cleveref}
\usepackage{colortbl}

\usepackage[toc,page]{appendix}

\newcommand*{\atanh}{{\tanh^{-1}}}

\renewcommand*{\epsilon}{\varepsilon}

\newcommand{\of}[1][t]{
	\mathopen{}\mathclose{}\bgroup\left(
	#1
	\aftergroup\egroup\right)
}

\begin{document}
	
	\title{Coherence properties of a spin in a squeezed resonator}
	\author{Inbar Shani}
	\author{Emanuele G. Dalla Torre}	
	\author{Michael Stern}	
	\affiliation{Department of Physics and Center for Quantum Entanglement Science and Technology, Bar-Ilan University, 52900, Ramat Gan Israel}
	\begin{abstract}		
A promising venue for hybrid quantum computation involves the strong coupling between impurity spins and superconducting resonators. One strategy to control and enhance this coupling is to prepare the resonator in a non-classical state, such as a squeezed state. In this work, we theoretically study the effects of these states on the coherence properties of the spin. We develop an analytic approach based on the Schrieffer-Wolf transformation that allows us to quantitatively predict the coupling and the dephasing rate of the spin, and we numerically confirm its validity. We find that squeezing can enhance the coupling between the resonator and the spin. However, at the same time, it amplifies the photon noise and enhances the spin decoherence. Our work demonstrates a major impediment in using squeezing to reach the strong-coupling limit.	
	\end{abstract}
	\maketitle
	\section{Introduction}	
	 Impurity spins in semiconductors are quantum entities with a long coherence time, which enables them to store safely quantum information \cite{muhonen2014storing}. Unfortunately, the weakness of their interaction
	 with the environment hinders our ability to control them directly. An appealing road towards a spin-based quantum processor consists of combining the impurity spins with superconducting circuits \cite{kurizki2015quantum,clerk2020hybrid}. 
	To realize this kind of hybrid system, one needs to reach the strong coupling regime, where the coupling between the spins and the superconducting circuit is much larger than the decoherence rates. In recent years, such a regime was reached for large ensemble of spins \cite{kubo2010strong,bienfait2016controlling}, and for spin-like macroscopic structures \cite{ viennot2015coherent, borjans2020resonant}. Yet, the coupling constant of a single microscopic spin with a superconducting resonator is extremely small (of the order of a few kHz \cite{haikka2017proposal}), and reaching the strong coupling, in these conditions, remains a great challenge \cite{lee2019ultrahigh}. Recently, Ref.~\cite{leroux2018enhancing} suggested to increase artificially the coupling by squeezing the resonator.  In a squeezed state, the fluctuations of the electromagnetic field in a given quadrature can be controlled and made arbitrarily large \cite{caves1981quantum, abadie2011gravitational,aasi2013enhanced}. Thus, increasing the coupling by squeezing seems interesting and even promising. On the other hand, the large number of photons that characterizes a squeezed state leads to large fluctuations, and may compromise the coherence of the spin. In this work, we study the interplay between these two effects and show that the beneficial effects of squeezing are strongly suppressed by the noise in the photon number. 
	\section{Physical model}
\begin{figure}[t]
		\centering
		\includegraphics[width=0.95\linewidth]{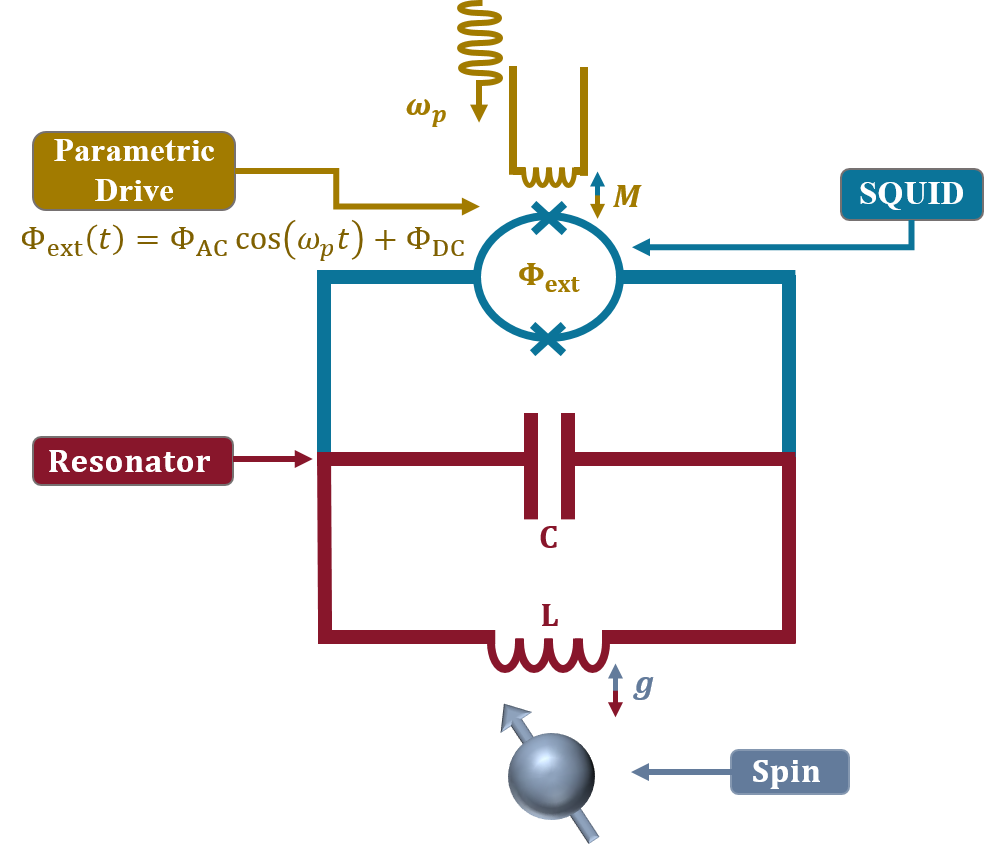}
		\caption{Circuit diagram showing a lumped element LC resonator (red) of resonance frequency $\omega_{r}$ and coupled inductively by coupling constant $g$ to a spin (gray). In order to increase the coupling $g$, one drives a SQUID (blue) with a parametric drive (yellow) at frequency $\omega_p=2\omega_{r}$. }
		\label{fig:proposal_scheme}
	\end{figure}

	We consider the quantum circuit illustrated in \cref{fig:proposal_scheme}, which contains a lumped element LC resonator of resonance frequency $\omega_{r}$, coupled inductively by a coupling constant $g$ to a spin of transition frequency $\omega_s$.	
	 As mentioned earlier, the coupling between the two systems is intrinsically small and it is necessary to increase it by at least one order of magnitude in order to reach the strong coupling regime. Following Ref.~\cite{leroux2018enhancing}, we explore the possibility of squeezing the resonator to enhance its coupling with the spin. To achieve this goal, we connect the resonator to a non-linear element, namely a superconducting quantum interference device (SQUID).
	The Hamiltonian of the system can be written as 
	\begin{align}
	\label{eq:hamiltonian_system_v1}
	\begin{split}
	H/\hbar&= \underbrace{\omega_{r} a^\dagger a}_{\text {LC resonator}}+\;\;\;\underbrace{\tfrac12\omega_{s}\sigma_z}_{\text{spin}}\; \;+\;\;\underbrace{ g(a+a^\dag)\sigma_x}_{\text{coupling}} \\&\;\;\;\;\;\;\;\;\;\;\;\underbrace{-2\frac{E_J}\hbar\cos\left(\frac{\Phi_{\rm ext}}{2\varphi_0}\right)\cos\left(\varphi\right)}_{\text{SQUID}}
	\end{split}
	\end{align}
	where $a^{\dagger}$ ($a$) is the creation (annihilation) operator of a photon in the resonator,  $\lbrace\sigma_z,\sigma_x\rbrace$ are the Pauli matrices, $E_J$ is the Josephson energy of the junctions, $\Phi_{\rm ext}$ is the flux threading the loop of the SQUID, $\varphi_0=\hbar/2e$ is the reduced flux quantum and $\varphi$ is the superconducting phase difference between the terminals of the inductor L \cite{girvin2011circuit}. The flux $\Phi_{\rm ext}$ is varied over time according to 
	\begin{align}
	\Phi_{\rm ext}= \Phi_{\rm DC}+\Phi_{\rm AC}\cos\left(\omega_pt\right).
	\end{align}
	If $\Phi_{\rm AC}\ll\varphi_0$,  it is possible to expand the cosine term in \cref{eq:hamiltonian_system_v1} to the first order around $ \Phi_{\rm DC}$, such that 
	\begin{align}
	\begin{split}
	\cos\left(\frac{\Phi_{\rm ext}}{2\varphi_0}\right) \approx\cos\left(\frac{\Phi_{\rm DC}}{2\varphi_0}\right)-\frac{\Phi_{\rm AC} }{2\varphi_0}\sin\left(\frac{\Phi_{\rm DC}}{2\varphi_0}\right)\cos (\omega_p t).
	\end{split}
	\end{align} 
	This leads to 
	\begin{align}
	\begin{split}
	H/\hbar&=\omega'_r a ^\dag a+\tfrac12\left(\omega'_r-\omega_{r}\right) \left(a^2+{a^\dag}^2\right)+\tfrac12\omega_s \sigma_z 
	\\&+g(a+a^\dagger)(\sigma_+ +\sigma_-)-\lambda
	\cos(\omega_p t) \left(a+{a^\dag}\right)^2,\end{split}
		\label{eq:system_hamitonian_v2}
	\end{align}
	where $\omega'_r =\omega_{r}+4\varphi_{\rm ZPF}^2\left( E_J/\hbar\right) \cos\left[\Phi_{\rm DC}/(2\varphi_0)\right]$ and
	\begin{align}
	\label{eq:Phisical_Lambda}\lambda = \frac{E_J}{\hbar}\Phi_{\rm AC} \;\frac{\varphi_{\rm ZPF}^2}{\varphi_0}\sin\left(\frac{\Phi_{\rm DC}}{2\varphi_0}\right).
	\end{align}
	We now move to a frame that rotates at the pump frequency, by applying the unitary transformation $U(t)=\exp\left[i\omega_p t/2 \left(a^\dagger a+\sigma_z/2\right)\right]$, leading to the transformed Hamiltonian $\tilde{H}=UHU^\dagger-i\hbar\dot{U}U^\dagger$. When $\omega_p \approx 2\omega'_r$, one can neglect the quickly rotating terms and obtain  
	\begin{align}
	\label{eq:squeezed_resonator_H}
	\begin{split}                       
	\tilde{H}/\hbar=&\tilde{\omega}_{r} a^\dagger a+\tfrac12\tilde{\omega}_{s}\sigma_z 
	- \tfrac{1}{2}\lambda \left(a^2+a^{\dagger2}\right)\\
	&+g \left(a\sigma_{+}+a^\dagger \sigma_{-}\right),
	\end{split}
	\end{align}
	where $\tilde{\omega}_{r}= {\omega'}_r -\omega_p/2$ and  $\tilde{\omega}_{s}=\omega_{s}-\omega_p/2$.  Eq.~(\ref{eq:squeezed_resonator_H}) describes the coupling between a spin and a squeezed resonator and is the focus of the present study.\\
	In order to characterize the effective coupling between the spin and the squeezed resonator, we diagonalize the latter using a Bogoliubov transformation. To perform this task, we introduce the canonical operators $\gamma$ and $\gamma^\dag$, defined as
	\begin{align}
	\begin{cases}
	\gamma =a \cosh(r) -a^\dag \sinh(r)\\
	\gamma^\dagger =a^\dag \cosh(r) -a \sinh(r)        
	\end{cases},
	\end{align}
	such that 
	\begin{align}
	\gamma^\dagger \gamma = \cosh(2r)a^\dag a-\tfrac12\sinh(2r)\left(a^2+{a^\dag}^2\right).
	\end{align}
	The Hamiltonian (\ref{eq:squeezed_resonator_H}) becomes
	\begin{align}
	\label{eq:squeezed_frame_hamiltonian}
	\begin{split}
	\tilde{H}/\hbar= & \Omega_r \gamma^\dagger \gamma +\tfrac12\tilde{\omega}_{s}\sigma_z+\tfrac12ge^r(\gamma^\dagger+\gamma)(\sigma_{+}+\sigma_{-}) \\
	& -\tfrac{1}{2}ge^{-r}(\gamma^\dagger-\gamma)(\sigma_{+}-\sigma_{-}).
	\end{split}
	\end{align}
	where $\Omega_r =\tilde{\omega}_{r}/{\cosh(2r)}$ and $r=\atanh(\lambda/\tilde{\omega}_{r})/2$. The Hamiltonian  \cref{eq:squeezed_frame_hamiltonian} describes two main effects of squeezing: First, the frequency of the resonator is reduced
	from $\tilde\omega_r$ to $\Omega_r$ and, second, the coupling between the spin and the resonator is enhanced by a factor of $e^r/2$. At first sight, this factor can be arbitrarily large and thus brings the system to the strong coupling regime \cite{leroux2018enhancing}.   However, as we will see in the following, this effect is impaired by the enhanced decoherence of the spin.
	
	To study the interplay between squeezing and decoherence, we consider the combined action of the Hamiltonian $\tilde{H}$ and the decay of photons from the resonator, described by the quantum master equation
	\begin{align}
	\label{eq:Lindblad_rho}
	\frac{d}{dt}\rho=-\frac{i}{\hbar}\left[\tilde{H},\rho \right] + L\rho L^\dagger -\tfrac12 \left(L^\dagger L\rho +\rho L^\dagger L\right),
	\end{align}
	where $\rho(t)$ is the density matrix and $L=\sqrt{\kappa}a $ is a Lindblad superoperator. The operator $L$ originates from the coupling between the superconducting resonator and the external environment, and drives the resonator to its vacuum (zero photons) state.
\section{Numeric simulation of a squeezed resonator}
\label{sec:numerics_of_squeezed_resonator}	
\begin{figure*}
	\includegraphics[width=0.75\linewidth]{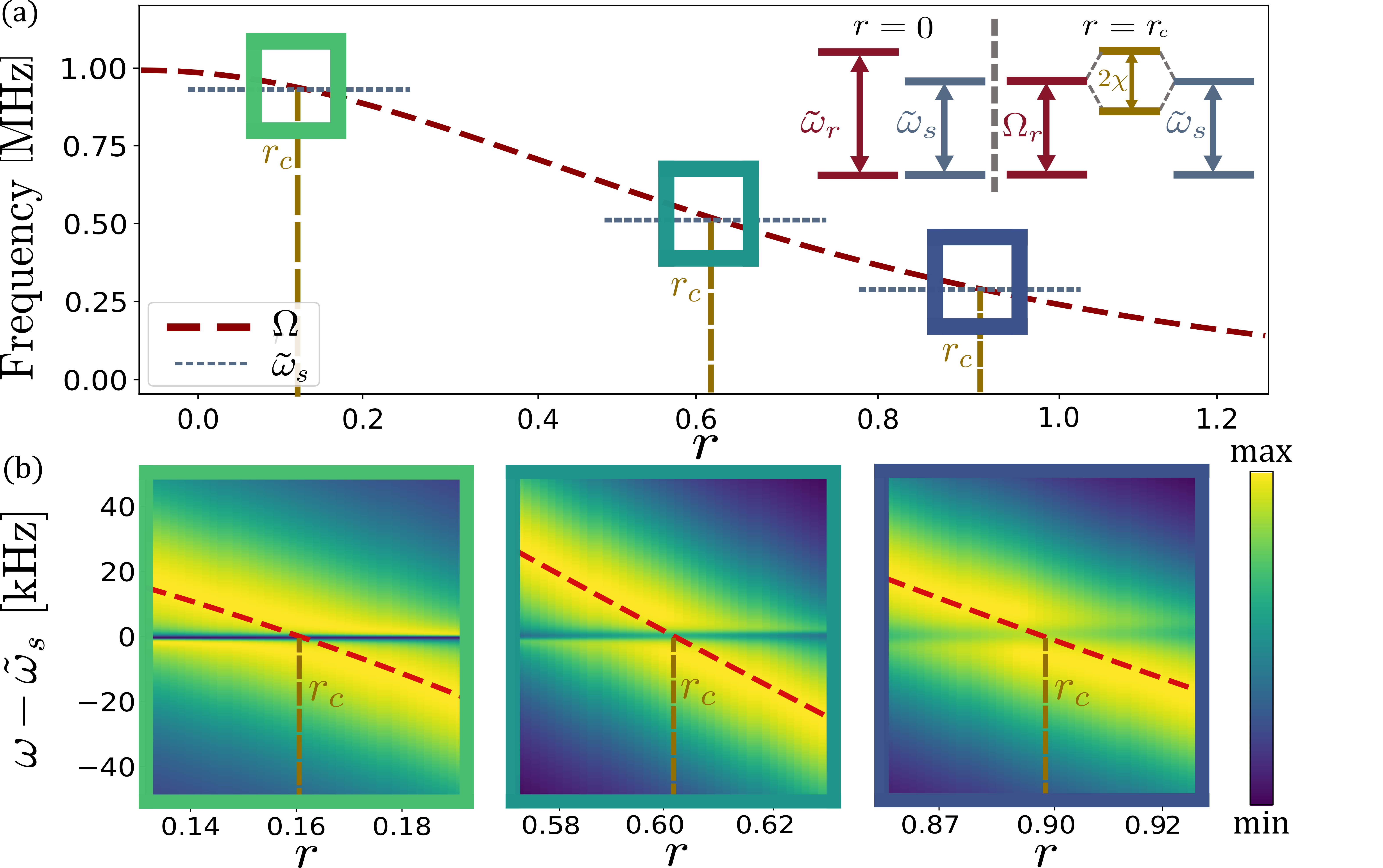}
	\caption{(a) Frequency of the squeezed resonator, $\Omega_{r}$, in the frame rotating at angular frequency $\omega_p/2$ relative to the laboratory frame, as a function of the squeezing parameter $r$. The resonator frequency is set to $\tilde\omega_r=1$ MHz and spin's frequency changes between $\tilde{\omega}_s= 0.9, 0.6, 0.3 $ MHz. Accordingly, the crossing point between the resonator and the spin changes between $r_c\approx0.16, 0.61, 0.89$. The inset shows a schematic picture of the energy levels, illustrating the resonance condition in the presence of finite squeezing. (b) Spectrum of the resonator $S_r(\omega)$, in the vicinity of the crossing conditions ($\Omega_r=\tilde{\omega}_s$) for $g=5$ kHz and $\kappa=40$ kHz.}
	\label{fig:anticross}
\end{figure*}	 
To quantify the coupling between the spin and the resonator, we numerically compute the power-spectrum of the resonator $S_r\left[\omega\right]$, defined as 
\begin{align}
\label{eq:power_spec}
S_r[\omega]=\int_{-\infty}^{\infty} \langle a^\dag(t)a(0) \rangle_\infty e^{-i\omega t } dt
\end{align}
where the sub-index $\infty$ implies that we compute the expression in the steady state \footnote{The numerical calculations presented in this work were obtained using the \textit{QuTiP} python package \cite{johansson2013qutip}. All the codes used to generate the figures in this article can be found online at \url{https://github.com/InbarShani2610/Coherence_Properties_of_a_Spin_in_a_Squeezed_Resonator}}. 
Since we work with ladder operators of an harmonic oscillator, and these cannot be described by a finite matrix, the precision of the numerical calculation depends on the truncation of the matrix operator which represents them. We truncate the matrix to a maximal number of photons, denoted by $N$, leading to density matrices of size $(2N)^2$. In the presence of squeezing, large values of $N$ are required to obtain results that coincide with  the exact solution (see \cref{fig:frames_comaprison} in \cref{ap:truncation_parameter}). 
To overcome this difficulty, we perform the numerical calculations in the squeezed frame, see \cref{eq:squeezed_frame_hamiltonian}, where smaller values of $N$ are sufficient to obtain good numerical results. Note that the Lindblad superoperator $L$ must transform accordingly:
\begin{align}
\label{eq:Lindblad_gamma}
L= \sqrt{\kappa}\cosh( r )\gamma + \sqrt{\kappa}\sinh( r )\gamma^\dag
\end{align}  

\begin{figure}[t]
	\centering	\includegraphics[width=0.95\linewidth]{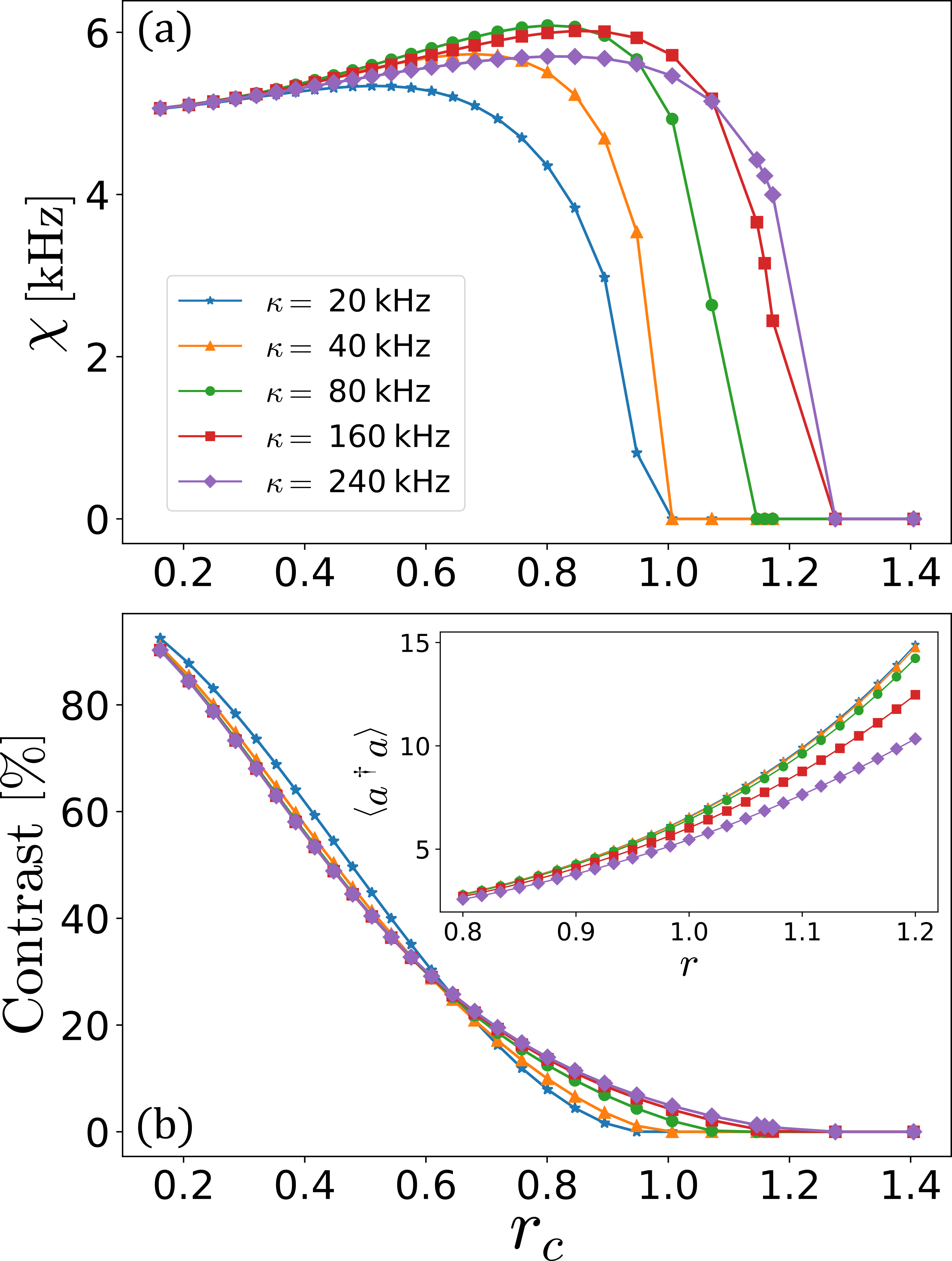}
	\caption{ (a) Dispersive shift $\chi$ determined from the spectrum of the resonator $S_r[\omega]$, as a function of $r_{c}$. (b) Contrast of the spectrum $1-S_{min}/S_{max}$. The suppression of the contrast at large squeezing indicates that enhanced decoherence of the spin. Inset: photon number occupying the resonator.}
	\label{fig:chi_contrast}
\end{figure}

Equation (\ref{eq:Lindblad_gamma}) represents a major difference between the present work and Ref.~\cite{leroux2018enhancing}, where the Lindblad operator was assumed to be proportional to the annihilation operator {\it in the squeezed frame}, namely $L'=\sqrt{\kappa}\gamma$. The physical realization of the Lindblad operator $L'$ requires one to squeeze the vacuum outside the resonator by exactly the same amount as the squeezing inside the resonator. While theoretically possible, this situation is unrealistic in an actual experiment. The substitution of $L'$ with $L$ has dramatic implications. In particular, this substitution leads to the disappearance of the level splitting shown in Fig. 1 of Ref. \cite{leroux2018enhancing} (see \cref{ap:frame_a_gamma}).\\ 
To study the effect of squeezing on the effective coupling between the resonator and the spin in a controlled manner, we consider a system where the bare frequency of the resonator is larger than the frequency of the spin $\tilde\omega_r>\tilde\omega_s$. For concreteness, throughout the article we consider a system with $\tilde\omega_r=1$ MHz, $\tilde\omega_s =0.6$ MHz, $g=5$ kHz, and $\kappa = 40$ kHz (unless explicitly mentioned otherwise). By introducing a squeezing term, we effectively reduce the resonator frequency to $\Omega_r$ (see \cref{eq:squeezed_frame_hamiltonian}), until the resonance condition is matched ($\Omega_{r}=\tilde{\omega}_s$), as illustrated in \cref{fig:anticross}(a). 
Changing the bare frequency of the spin $\tilde{ \omega}_s$ modifies the value of the squeezing parameter $r=r_c$ at resonance condition. The power spectrum near $r_{c}$, \cref{fig:anticross}(b),  has the typical structure of an avoided level-crossing. At a fixed value of $r$, the spectrum shows two peaks, whose frequencies correspond to the energy levels of the mixed resonator-spin states (see inset in \cref{fig:anticross}(a)).
We use the frequency difference between the maximum and the local minimum of the spectrum at resonance, i.e. for $r=r_c$, to estimate the coupling strength between the spin and the resonator. In analogy to the Jaynes-Cummings model, we denote this distance as $\chi$ and plot its value as a function of $r_c$ in \cref{fig:chi_contrast}(a). For small bare detunings ($\left|\tilde{ \omega}_r-\tilde{ \omega}_s\right|\ll\tilde{ \omega}_r$), the size of the anticrossing $\chi$ increases as a function of $r_c$ until it reaches a maximal value. This result is in stark contrast to the case of the squeezed vacuum operators of Ref. \cite{leroux2018enhancing}, where arbitrarily large couplings can be obtained.

The upper limit of $\chi$ is due to the back-action of the squeezed resonator on the spin, leading to its fast dephasing: this process can be identified by observing the broadening of the spectrum at resonance (see \cref{fig:anticross}(b)). We quantify this effect by measuring the ratio between the maximal intensity of the spectrum. We denote as $S_{\rm min}$ (resp. $S_{\rm max}$) the values of the spectrum at its local minimum (resp. maximum) and show in \cref{fig:chi_contrast}(b) that the contrast of the anticrossing, defined as $1-S_{\rm min}/S_{\rm max}$, is a monotonously decreasing function of $r_c$. When the resonator is squeezed significantly ($r_c\sim1$) the contrast vanishes, indicating that the spin is completely dephased. To analyze the physical origin of this effect, we repeat the same calculations for different values of the loss rate $\kappa$, which controls the number of photons occupying the resonator (see the inset of \cref{fig:chi_contrast}(b)). The maximum size of the anticrossing $\chi$ is shifted towards higher values when the decay rate of the resonator $\kappa$ is increased. Eventually, for large values of the decay rate $\kappa$, $\chi$ begins to deteriorate. The maximal value of $\chi$ is only 20$\%$ larger than its initial value and is insufficient to reach the strong coupling regime. 
\section{Schrieffer-Wolff Transformation for a Squeezed resonator }
Motivated by the numerical results of the previous section, we now introduce a systematic approach to describe the effects of the squeezed resonator on the spin. Our approach generalizes the Schrieffer-Wolf (SW) transformation \cite{schrieffer1966relation} to a squeezed resonator and allow us to compute the energy shift and the decoherence of the spin analytically. As a first step, we write the Hamiltonian (\ref{eq:squeezed_resonator_H}) as $\tilde{H}=\tilde{H}_0+V$, with
\begin{align}
\label{eq:H0tilde}
\tilde{H}_0/\hbar&= \tilde{\omega}_{r} a^\dag a + \tfrac12 \tilde{\omega}_{s} \sigma_{z}-\tfrac12 \lambda\left(a^2+{a^\dagger}^2\right)\\
\label{eq:V_coupling}
V/\hbar&= g(a^\dag\sigma_-+a\sigma_+).
\end{align}
Next, we apply the unitary transformation $U=e^{S}$, where $S$ is the anti-Hermitian operator
\begin{align}
\label{eq:SW_transform}
S &= C_1 (a\sigma_{+}-a^\dag \sigma_-)+C_2 (a \sigma_- - a^\dag\sigma_{+}),  ~{\rm with}~ \\
C_1 &= \frac{g }{\Delta} \frac{
	\tilde{\omega}_{s}+	\tilde{\omega}_{r}}{\tilde{\omega}_s+\Omega_r},~~~
C_2 = \frac{g  }{\Delta}\frac{\lambda}{\tilde{\omega}_s+\Omega_r}
\end{align}
and $\Delta=\tilde{\omega}_{s}-\Omega_r$. This transformation satisfies the condition $[\tilde{H}_0,S]=V$  and leads to a transformed Hamiltonian $\tilde{H}_{\rm SW}=e^S \tilde{H} e^{-S}$ that does not include linear terms in $V$. By expanding this expression to second order in $V$, we obtain  $\tilde{H}_{\rm SW}/\hbar=  \tilde{H}_0/\hbar+\delta\tilde{\omega}_s \sigma_z/2$, with a dispersive shift
\begin{align} 
\label{eq:dws1} \delta\tilde{\omega}_s&=\chi_r (2\tilde{n}+1),
\end{align}
where we defined $\chi_r={g}^2 /\Delta$, and
\begin{align}
\label{eq:NplusHalf}  
\tilde{n}+\tfrac12=\frac{\left(\tilde{\omega}_{s}+\tilde{\omega}_{r}\right)\left( 2 a ^\dag a+1 \right) -\lambda \left( a^2+{a^\dag}^2\right)}{2(\tilde{\omega}_s+\Omega_r)}
\end{align}
For $\lambda=0$, \cref{eq:dws1,eq:NplusHalf} recover the well known Lamb shift of a spin coupled to a resonator
\begin{align}
\delta\tilde{\omega}_s=& \chi_0\left(2a^\dag a +1\right)~~~ {\rm and}~~~ \chi_0 =\frac{g^2}{\tilde{ \omega}_s - \tilde{ \omega}_r}.
\label{eq:shift_thermal}
\end{align}

\paragraph*{}The SW transformation creates an effective channel of dissipation for the spin, in addition to its intrinsic dissipation. In order to consider this effect, one needs to transform the dissipation operator $L$ using the SW transformation. Up to second order in $V$, 
	\begin{align}
	\begin{split}
	&\tilde{L}_{\rm SW}=e^{S} L  e^{-S}
	= L +\left[S,L\right] +\tfrac{1}{2}\left[S,\left[S,L\right]\right]+ ...
	\end{split}
	\end{align}
	Using \cref{eq:SW_transform}, we obtain
	\begin{align}
	\label{eq:LSW}
	L_{\rm SW}=\sqrt{\kappa}a+\sqrt{\kappa} C_1 \sigma_- +\sqrt{\kappa} C_2\sigma_+  +{\rm O}\left(\frac{g^2}{\Delta^2} \right)
	\end{align}
	After the transformation, the resonator and the spin are decoupled, such that the coherence of the spin is entirely dictated by the second and third terms of \cref{eq:LSW}. If we consider these two terms as independent channels, we obtain an effective decoherence rate
	\begin{align}
		\label{eq:gamma_purcell}
		\begin{split}
		\Gamma_{\rm Purcell}&=\Gamma_1+\Gamma_2=\kappa\left(C_1^2+C_2^2\right).
		\end{split}
	\end{align}
	In the limit of $\lambda=0$, this expression reproduces the well-known Purcell decay, $\Gamma_{\rm Purcell}=g^2 \kappa/\Delta^2$ \cite{purcell1995spontaneous,bienfait2016controlling}. 
\section{Photon noise}	
In addition to the Purcell effect, the squeezing of the resonator also affects the pure dephasing rate of the spin. Stochastic fluctuations in the number of photons in the resonator create a random dispersive shift which translates into a dephasing of the spin. The rate of this process can be determined by the decay of the expectation value  $\langle\sigma_x(t)\rangle$ in a system initialized in the $|+\rangle=\left(|0\rangle+|1\rangle\right)/\sqrt{2}$ state \cite{ithier2005decoherence},
	\begin{align}
    \langle\sigma_x (t)\rangle =\frac{1}{2}\left(
	e^{i\int_{0}^{t}\delta\tilde{ \omega}_s (t')dt'}
	+e^{-i\int_{0}^{t}\delta\tilde{ \omega}_s (t')dt'} \right).
	\end{align}
	Assuming that $\tilde{n}(t)=\tilde{n}_0+\delta \tilde{n}{(t)}$, where the fluctuation $\delta \tilde n(t)$ is a random variable with zero average, one can rewrite the exponent as 
 	\begin{align}
 	e^{\pm i\int_{0}^{t}\delta\tilde{ \omega}_s (t')dt'}=e^{\pm i\chi_r\left(2{\tilde{n}_0}+1\right) t}e^{\pm i2\chi_r\int_{0}^{t}\delta \tilde{n}(t')dt'}.
 	\end{align}
	By expanding the exponent to a Taylor series, one obtains
		\begin{align}
		\langle e^{\pm i\chi_r\int_{0}^{t} \delta \tilde{n}(t')dt'}\rangle&\approx  1 -2\chi_r^2\int_{0}^{t}dt'\int_{0}^{t}dt''~\langle \delta \tilde{n}(t') \delta \tilde{n}(t'')\rangle\nonumber \\
		&\approx 1 - 2\chi_r^2 t \int_{-\infty}^\infty \langle \delta \tilde n(\tau)\delta \tilde n (0)\rangle_{\infty}d\tau.	\end{align}
		Here, the last identity is valid in the limit of $t\to\infty$, under the assumption that the resonator is found in a steady state, where two-time correlations depend on the time difference only. In the case of Gaussian fluctuations, the higher order terms can be re-summed exactly leading to 
	\begin{align}
	\langle\sigma_x(t)\rangle =& \cos\left[\chi_r\left({2\tilde{n}_0}+1\right)t\right] e^{-\Gamma_\phi^{\rm photon} t},
	\end{align}	where we defined
	\begin{align} 
	\label{eq:gamma_photon}
	&\Gamma_\phi^{\rm photon}=2\chi_r^2\tilde\eta,~{\rm and}~ \tilde\eta = \int_{-\infty}^{\infty} \langle \delta \tilde{n}(\tau) \delta \tilde{n}(0)\rangle d\tau.
	\end{align}
	In \cref{ap:density_correlations}, we compute the correlations $\langle \delta n(\tau)\delta n(0) \rangle$ for a thermal state and for a coherent state and reproduce known results for the photon noise dephasing of the spin.
\section{Steady state correlations of a squeezed resonator}
In the previous sections we expressed the dispersive shift and the decoherence rates of the spin in terms of physical observables of the resonator, see  \cref{eq:dws1,eq:gamma_purcell,eq:gamma_photon} . In the following, we compute these quantities in the steady state of a squeezed resonator described by the Hamiltonian (\ref{eq:H0tilde}) (with $g=0$).	
According to the Lindblad master equation (\ref{eq:Lindblad_rho}), the expectation value of a generic operator $O(t)$ is determined by the differential equation
\begin{align}
\begin{split}
\frac{d}{d t}\left<O(t)\right>&=\frac{i}{\hbar}\left<\left[H,O(t)\right]\right> +\left<L^\dagger O(t) L\right> 
\\
& -\tfrac12\left(\left< L^\dagger L O(t) \right> +\left< O(t) L^\dagger L \right> \right).\label{eq:masterO}
\end{split}
\end{align}
This equation can be explicitly solved if one finds a set of variables $\vec{O}=(O_1, O_2, ...)$ whose expectation values satisfy the closed-form recursive relation
\begin{align}
\label{eq:EOM_1op}
 \frac{d}{d t}\left<O_{i}(t)\right> = \sum_j G_{i,j}\left< 	O_{j}(t)\right>,
 \end{align}
where $G_{i,j}$ is a time independent scalar. For a squeezed resonator, this condition is satisfied by the operator $\vec{O} =  \begin{pmatrix} a^\dagger a, \ a^{2}, \ {a^{\dagger}}^{2}, \ \mathds{1} \end{pmatrix}^{T}$ with
	\begin{align}
	\label{eq:EOM_mat}
	G=\left(\begin{matrix}
	-\kappa      & -i \lambda         & i \lambda          & 0         \\
	i2  \lambda  & -i2\tilde{\omega}_{r}-\kappa & 0                  & i\lambda  \\
	-i2  \lambda & 0                  & i2\tilde{\omega}_{r}-\kappa & -i\lambda \\
	0            & 0                  & 0                  & 0
	\end{matrix}\right).
	\end{align}
The steady-state expectation values of $\vec{O}$ can be found by equating the left-hand side of \cref{eq:EOM_1op} to 0, or equivalently by demanding that $ \sum_j G_{i,j}\left<O_{j}(t)\right>_\infty=0$. Using this approach, we find
	\begin{align}
	\langle a^\dagger a  \rangle_\infty &= \frac{2 \lambda ^2}{4 \Omega_r^2+\kappa ^2}\\
	\langle  a^2  \rangle_\infty &= \frac{\lambda  (2 \tilde{\omega}_r+i \kappa )}{4 \Omega_r^2+\kappa ^2}\\
	\langle  {a^\dag}^2  \rangle_\infty& = \frac{\lambda  (2 \tilde{\omega}_r-i \kappa )}{4 \Omega_r^2+\kappa ^2}.
	\end{align}
Plugging these results in \cref{eq:dws1}, we obtain an analytic expression for the dispersive shift of the spin,
\begin{align}
\label{eq:dis_shift}
\delta\tilde{\omega}_s = \frac{g^2 }{\tilde{ \omega}_s^2-\Omega_r^2}\left(     \frac{4 \lambda ^2\tilde{ \omega}_s}{4 \Omega_r^2+\kappa ^2}+\tilde{ \omega}_r+\tilde{ \omega}_s	\right)
\end{align}
\begin{figure}
	\centering
	\includegraphics[width=1\linewidth]{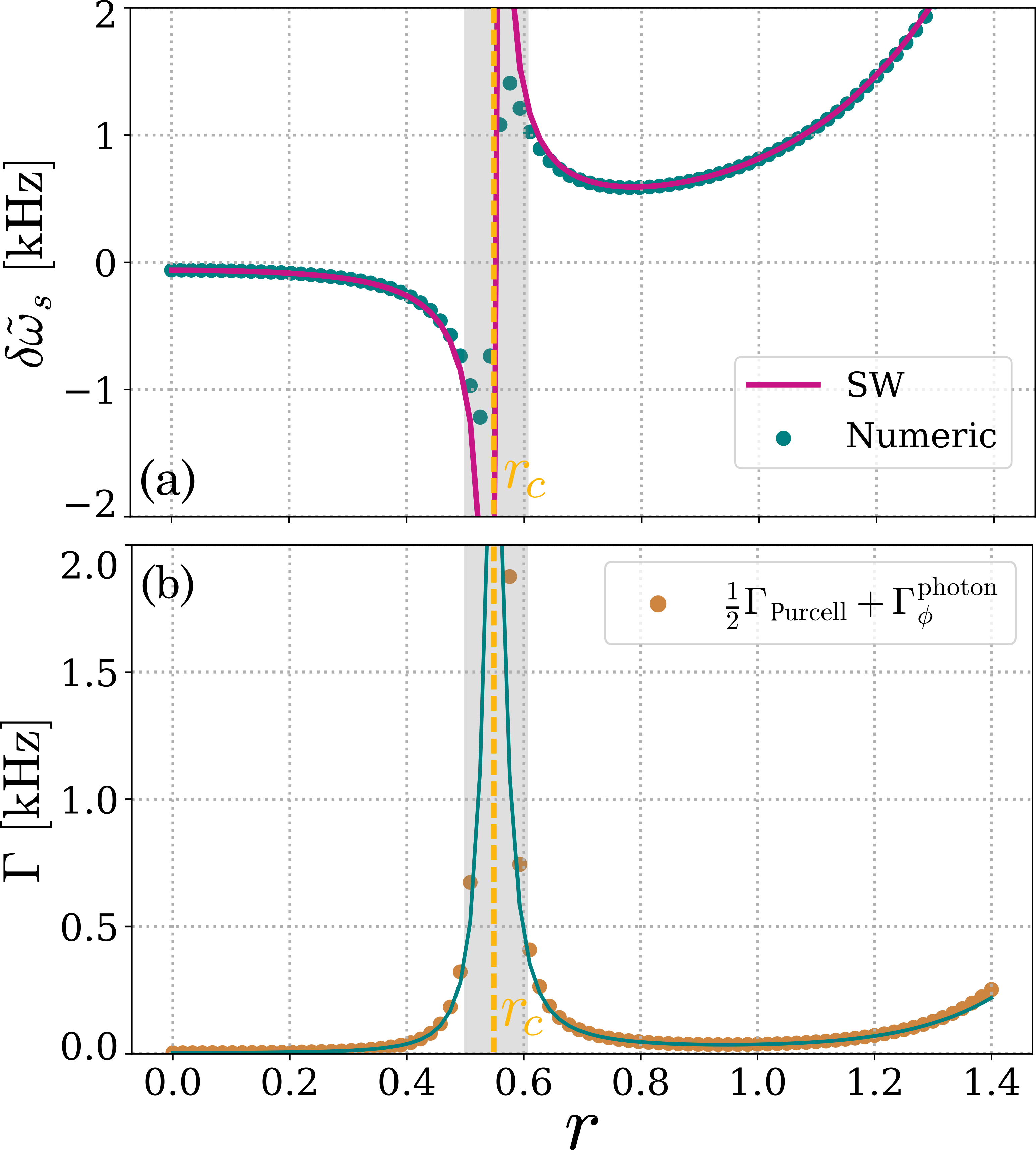}
	\caption{(a) Dispersive shift and (b) dissipation rate of the spin, as a function of the squeezing parameter $r$. The continuous lines are our analytic expressions for the dispersive shift, \cref{eq:dis_shift}, and for the total dissipation $\Gamma_{\rm tot}=\Gamma_{\rm Purcell}/2+\Gamma_\phi^{\rm photon}$,  \cref{eq:gamma_purcell,eq:squeezed_gamma_photon}. This approximation fits very well to the exact numerical results (dots), except for the range$|\Delta| \ll g$ (the gray shaded area marks where $|\Delta|<0.02g$). }
	\label{fig:Dispersive_shift_width_spin}
\end{figure}
For a quantum system described by a Markovian master equation, the two-time correlator of \cref{eq:gamma_photon} can be computed using the quantum regression theorem \cite{lax1967quantum} (see also \cref{ap:QRT} for a simple proof). According to this theorem, given a set of operators $\vec{O}$ that satisfy Eq.~(\ref{eq:EOM_1op}), the two-time correlation functions satisfy the differential equation
\begin{figure*}
	\centering
	\begin{tabular}{m{0.4\linewidth} m{0.4\linewidth}}
		\begin{center}(a) Numerics\end{center} & \begin{center}(b)  Analytic\end{center}
	\end{tabular}
	\includegraphics[width=0.85\linewidth]{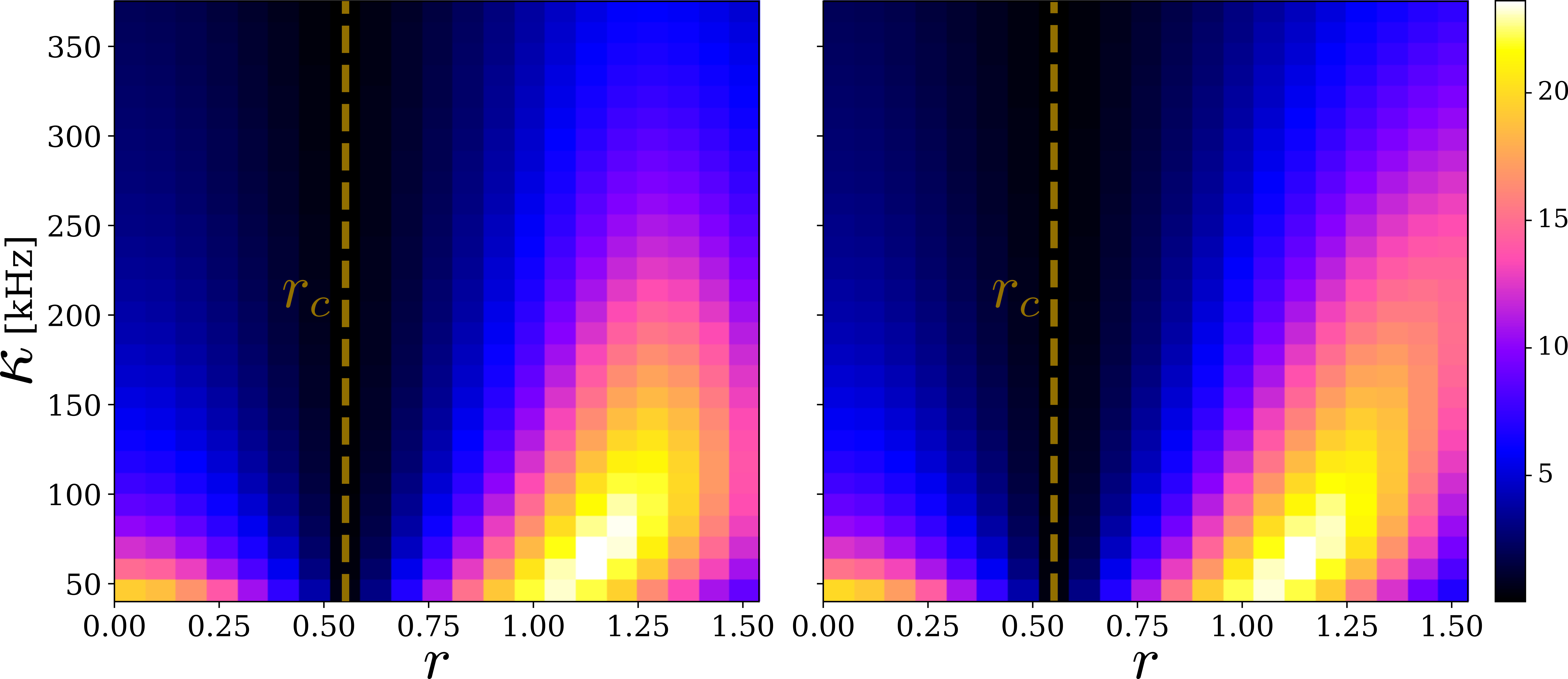}
	\caption{Ratio between the dispersive shift of a spin and its spectral width,  $\delta\tilde\omega_s/\Gamma$, as a function or the squeezing parameter $r$ and the loss-rate of the resonator $\kappa$. The ratio has a local maximum for $r = 1.1\pm 0.1$, where one obtains an optimal value for the effective coupling $\chi$, see Fig.~\ref{fig:chi_contrast}.}
	\label{fig:spin_spec_ratio}
\end{figure*}
\begin{align}
\label{eq:twotime_corr}
\frac{d}{d \tau}\left< 	O_{i}(t+\tau)O_{k}(t)\right> = \sum_j G_{i,j}\left< 	O_{j}(t+\tau)O_{k}(t)\right>.
\end{align} 
Using this theorem, we can compute the two-time correlation function of the squeezed resonator (see \cref{ap:density_correlations} for details). Plugging this result  into \cref{eq:gamma_photon} we obtain the dephasing rate due to photon noise in the squeezed resonator:
\begin{align}
\label{eq:squeezed_gamma_photon}
\Gamma_{\phi}^{\rm photon}=&\frac{2\chi_r^2\lambda^2}{\kappa\left(\Omega_r+\tilde{ \omega}_s\right)^2}  \left[\frac{ \left(2\Omega_r^2 +\kappa ^2
	\right) }{ \left(4 \Omega_r^2+\kappa ^2\right)}
+
4 \tilde{\omega}_{r} \tilde{\omega}_{s}\frac{  \left(4 \Omega_r^2+3 \kappa ^2\right)    }{  \left(4 \Omega_r^2+\kappa ^2\right)^2}   \right. \nonumber\\
&\left.  +	
2 \tilde{\omega}_{s}^2 \frac{\left(4 \tilde{\omega}_{r}^2+\kappa ^2\right) \left(4 \Omega_r^2+5 \kappa ^2\right)}{  \left(4 \Omega_r^2+\kappa ^2\right)^3}\right]. 
\end{align}
Equation~(\cref{eq:squeezed_gamma_photon})  is proportional to $\lambda^2$: The photon noise is associated with the occupation of the resonator, induced by squeezing, and vanishes for $\lambda=0$. See also  \cref{ap:density_correlations} for the cases of a resonator in a coherent state and in a thermal state. In all cases, the resulting expression is inversely proportional to $\kappa$: if the photons are allowed to escape rapidly from the resonator, they have a smaller effect on the dephasing of the spin.
\section{Numeric simulation of a spin in a squeezed resonator}
To test the validity of the expressions obtained for the dispersive shift of a spin, \cref{eq:dis_shift}, and the dissipation rate of the spin, \cref{eq:squeezed_gamma_photon,eq:gamma_purcell}, we compute numerically the power spectrum of of the spin $S_s\left[\omega\right]$, defined as
\begin{align}
\label{eq:power_spec_spin}
S_s[\omega]=\int_{-\infty}^{\infty} \langle \sigma_{x}(\tau)\sigma_{x}(0) \rangle_\infty e^{-i\omega \tau } d\tau
\end{align}
The dispersive shift is the distance between the position of the maximum of $S_s\left[\omega\right]$ and the bare frequency $\tilde{\omega}_s$. This quantity is plotted in \cref{fig:Dispersive_shift_width_spin}(a) as a function of $r$. The dispersive shift of the spin is initially negative (because $\tilde\omega_s<\tilde\omega_r$) and its absolute value increases as a function of $r$ for all $r<r_c$. At $r=r_c$, the dispersive shift changes sign and starts to decrease. At large $r\gtrsim 1$, the dispersive shift increases again, signaling an enhancement of the coupling between the spin and the resonator. Our numerical findings are in excellent agreement with our analytical result, Eq.~(\ref{eq:dis_shift}) , except for a narrow region around $r\approx r_c$, where the analytical curve diverges, while the numerical one remains finite. From the width of the numerical spectrum, we estimate the dephasing rate $\Gamma$, see \cref{fig:Dispersive_shift_width_spin}(b). We find a quantitative agreement with the analytical prediction $\Gamma = \Gamma_{\rm Purcell}/2+\Gamma_\phi^{\rm photon}$, where $\Gamma_{\rm Purcell}$ and $\Gamma_\phi^{\rm photon}$ are given in \cref{eq:squeezed_gamma_photon,eq:gamma_purcell}.\\

From the computed values of the dispersive shift and decoherence rate, we can estimate the relative strength of these two effects. The ratio between these two quantities is shown in Fig.~\ref{fig:spin_spec_ratio} and, again, shows an excellent agreement between analytics and numerics. At a fixed $\kappa$, this ratio follows a non-monotonic behavior, with local minima at resonance, for $r\approx r_c$, and at large squeezing, for $r\gg 1$. In these regions, the ratio between the dispersive shift and the decoherence rate tends to zero, indicating that the dephasing effect of squeezing dominates over the enhanced coupling between the spin and the resonator. This ratio obtains an optimal value for $r\approx 1.1$, in correspondence to the local maximum of the effective coupling parameter $\chi$ found in \cref{fig:chi_contrast}(a).
\section{Discussion and Conclusion}
In conclusion, we studied the coherence properties of a spin embedded in a resonator under parametric drive. We found out that such a drive gives rise to an enhanced effective coupling between the spin and the resonator but also increases relaxation and decoherence rates of both systems.  These non-unitary effects are related to the relatively large number of photons in the squeezed resonator. In particular, the number of photons in a squeezed resonator modifies the relaxation rate of the spin. This effect limits the relative enhancement of the coupling by squeezing to approximately $20\%$. This situation is in contrast to what usually occurs for coherent or thermal states, where the relaxation of the spin via the resonator - the so-called Purcell rate - is fixed by the intrinsic properties of the system and is independent of the photon occupation. Here, the squeezing terms change both the coupling between the two systems and their loss rates. The theoretical methods developed in this work enabled us to consider these two effects on equal footing and could be extended further to more ‘exotic’ quantum states like cat states \cite{vlastakis2013deterministically} or Gottesman-Kitaev-Preskill (GKP) states \cite{gottesman2001encoding,campagne2020quantum}. Our analytical approach based on generalized Schrieffer Wolf transformations and quantum regression theorem are valid for squeezed cavities on a large parameter scale and were found to be in quantitative agreement with the numerical calculations.
\begin{acknowledgements}
This work was supported by the Israel Science Foundation, Grants No. 426/15, 151/19, 154/19 and 898/19. I. S. acknowledges support from the Institute for Nanotechnology and Advanced Materials at Bar-Ilan University. We wish to thank T. Kontos, D. Vion and L. Bello for inspiring and fruitful discussions.
\end{acknowledgements}
	\appendix
	\section*{Appendix}
	\section{Comparison of the two decay mechanisms}
	\label{ap:frame_a_gamma}
		\begin{figure}[t]
	\includegraphics[width=1\linewidth]{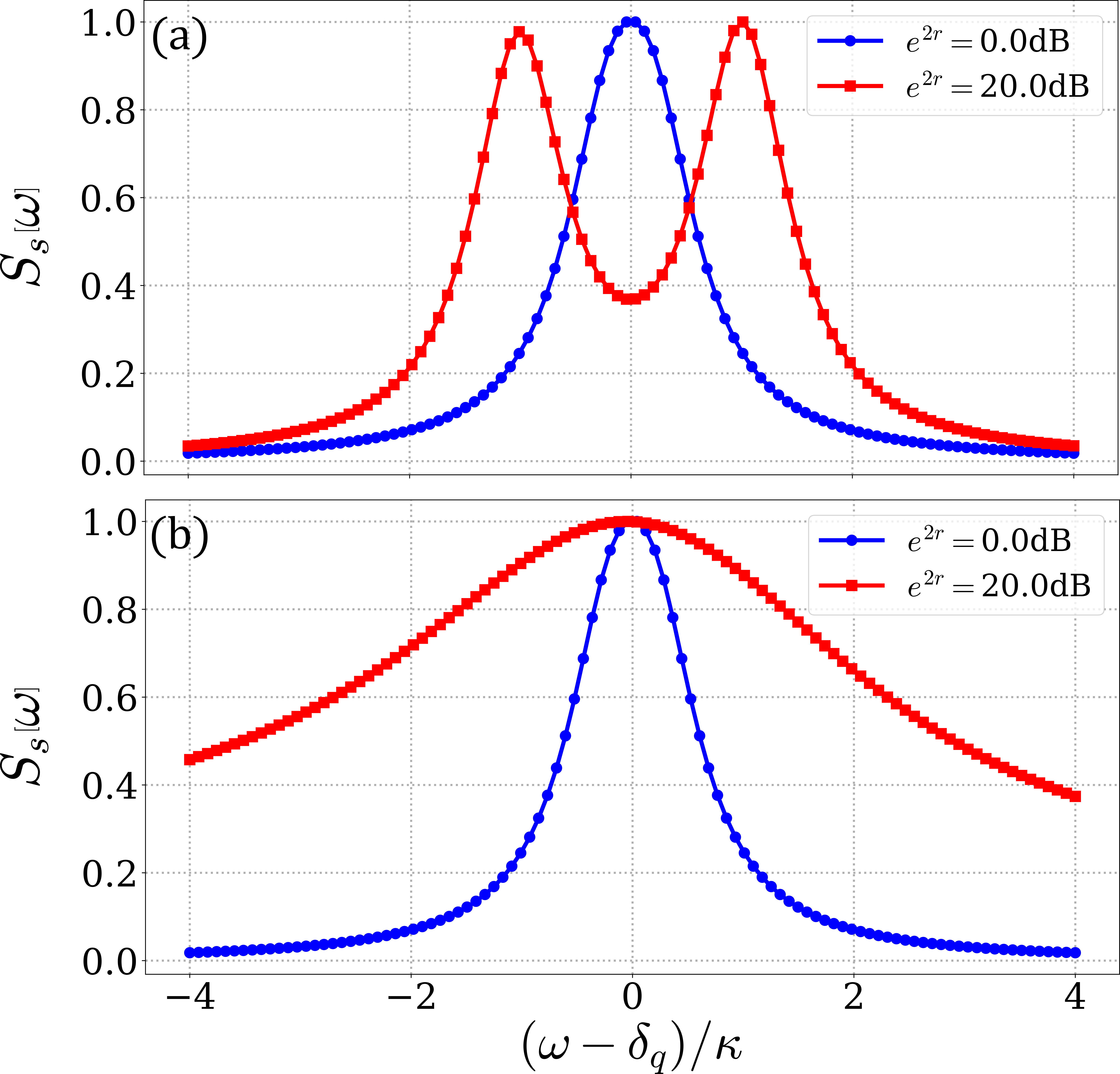}
	\caption{Spectrum of the spin, in the absence of squeezing (blue) and in the presence of squeezing (red). The upper and lower panel differ in the nature of the Lindblad operator used in the calculation: in (a) we use $L'=\sqrt{\kappa}\gamma$, which corresponds to a resonator that is immersed in a squeezed vacuum. In (b) we use $L=\sqrt{\kappa}a$, which corresponds to the physical situation of an unsqueezed vacuum. In the presence of squeezing, a splitting is observed for (a) but not for (b). }
	\label{fig:compare_a_gamma}
\end{figure}	
In this section we compare the effects of Lindblad operators coupled to the physical operator $a$, or to the squeezed resonator operator $\gamma$. \cref{fig:compare_a_gamma} shows the spectrum of the spin, defined in \cref{eq:power_spec_spin}, for the two cases. \cref{fig:compare_a_gamma}(a) reproduces Fig. 1 of Ref. \cite{leroux2018enhancing} and shows that in the presence of squeezing, the system reaches the strong coupling regime. \cref{fig:compare_a_gamma}(b) describes the physical situation, where the resonator is immersed in the regular vacuum. In this case, no level splitting is observed.	

	\section{Numerical calibration of the truncation parameter $N$}
	\label{ap:truncation_parameter}

In this appendix we study the effect of the truncation parameter $N$, and compare two different frames. The first frame, which we denote as the original frame, corresponds to the Hamiltonian in \cref{eq:squeezed_resonator_H} and $L=\sqrt{\kappa}a$. The second frame, which we denote as the squeezed frame, corresponds to the transformed Hamiltonian \cref{eq:squeezed_frame_hamiltonian} and Lindblad operator in the form of \cref{eq:Lindblad_gamma}. In each frame, we truncate the matrix representing the annihilation and creation operators, respectively $a,a^\dag$ and $\gamma,\gamma^\dag$, and vary the maximal number of photon occupation $N$.

\begin{figure}
	\centering
	\includegraphics[width=0.9\linewidth]{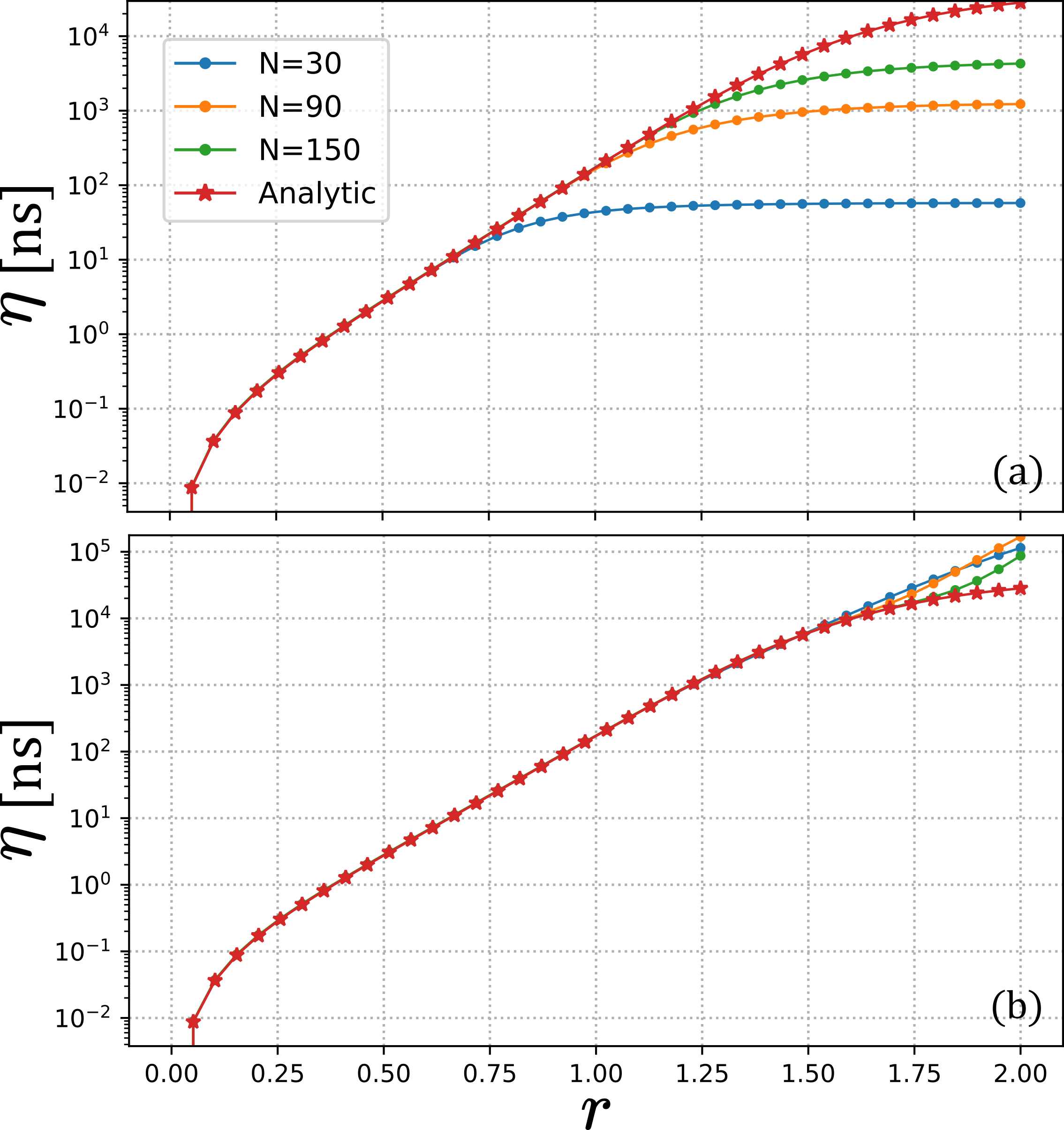}
	\caption{Integrated two-time correlation function $\eta$, defined in \cref{eq:eta}, for different values of the numerical truncation parameter $N$, (a) in the original frame and (b) in the squeezed frame.}
	\label{fig:frames_comaprison}
\end{figure}

For concreteness, we consider the integrated two-time correlation 
\begin{align}\eta = \int_{-\infty}^\infty \langle (n(\tau)-\bar{n})(n(0)-\bar{n})\rangle d\tau
\end{align}
for a squeezed resonator, whose analytical expression is computed in \cref{ap:density_correlations}, \cref{eq:eta}. In the original frame, \cref{fig:frames_comaprison}(a), the results converge slowly to the analytic solution as we increase $N$. For intermediate amount of squeezing, $r\approx 1.5$, the numerical solution requires extremely large values of the truncation parameter $N\gtrsim 150$ to converge to the analytic solution. In contrast, the numerical results in the squeezed frame, \cref{fig:frames_comaprison}(b), are less sensitive to the truncation of the matrix size, thus converging much faster to the analytic result. In this article we are interested in squeezing parameters $r\lesssim 1.5$ and, hence, we work in the squeezed frame and use $N=30$.

\section{Quantum Regression Theorem}
	\label{ap:QRT}
	 In this appendix we offer a simple derivation of the quantum regression theorem for Markovian master equations of the form
	\begin{align}
	\label{eq:rho_dot}
	\frac{d}{dt}\rho =\mathcal{L}\rho,
	\end{align}
	where $\mathcal{L}$ is the Liouvillian superoperator and $\rho$ is the density matrix. For time-independent $\mathcal{L}$, the time evolution of $\rho$ is given by 
	\begin{align}
	\rho(t+\tau)=e^{\mathcal{L}\tau}\rho(t).
	\end{align}
	The expectation value of an operator $A$ is defined as
		\begin{align}
	\langle A(t) \rangle= {\rm Tr}\left[A\rho(t)\right].
	\end{align}
	Its time derivative is given by
	\begin{align}
	\frac{d}{dt}\langle A(t)\rangle=  \frac{d}{dt}{\rm Tr}\left[A\rho(t)\right]= {\rm Tr}\left[A \frac{d}{dt}\rho(t)\right].
	\end{align}
	Plugging-in \cref{eq:rho_dot} we obtain
	\begin{align}
	\frac{d}{dt}\langle A(t)\rangle=   {\rm Tr}\left[A\mathcal L\rho(t)\right].
	\label{eq:dAdt2}
	\end{align}	
	The quantum regression theorem allows one to compute the two-time correlation function
	\begin{align}
	\label{eq:corr_2t}
	\langle A(t+\tau)O(t)\rangle = {\rm Tr}\left[ A e^{\mathcal{L}\tau}O\rho(t)\right]
	\end{align}
	for systems where the time derivative of $d \langle A(t)\rangle/dt$ has a linear dependence on the system's operators $B_j$,
	\begin{align}
	\label{eq:A_dot}
	\frac{d}{dt}\langle A(t)\rangle=\sum_j G_j \langle B_j(t)\rangle.
	\end{align}
	Because the trace of sum of matrices is the sum of their traces, we can rewrite \cref{eq:A_dot} as
	\begin{align}
	\frac{d}{dt}\langle A(t)\rangle=\sum_j G_j {\rm Tr}\left[ B_j \rho(t)\right] = \sum_j  {\rm Tr}\left[ G_jB_j \rho(t)\right]
	\label{eq:dAdt}
	\end{align}
	Because Eq.~(\ref{eq:dAdt}) and Eq.~(\ref{eq:dAdt2}) are satisfied for any $\rho(t)$, we obtain that
	\begin{align}
	\label{eq:AL_to_GB}
	A\mathcal{L}=\sum_j G_j B_j 
	\end{align}
	This result allows us to compute the time derivative of \cref{eq:corr_2t},
	\begin{align}
	\frac{d}{d\tau}\langle A(t+\tau)O(t)\rangle & =\frac{d}{d\tau} {\rm Tr}\left[ A e^{\mathcal{L}\tau}O\rho(t)\right] \\
	&= {\rm Tr}\left[ A \mathcal{L}e^{\mathcal{L}\tau}O\rho(t)\right].
	\end{align}
	Using \cref{eq:AL_to_GB} we obtain
	\begin{align}
	\frac{d}{d\tau}\langle A(t+\tau)O(t)\rangle & = {\rm Tr}\left[ \sum_j G_j B_j e^{\mathcal{L}\tau}O\rho(t)\right]\\
	&= \sum_j G_j {\rm Tr}\left[  B_j e^{\mathcal{L}\tau}O\rho(t)\right]\\
	& = \sum_j G_j \langle B_j(t+\tau)O(t)\rangle
	\label{eq:QRT2}
	\end{align}
	Eq.~(\ref{eq:QRT2}) is the quantum regression theorem used in Eq.~(\ref{eq:twotime_corr}).	
\section{Two-time correlation functions of the resonator}
\label{ap:density_correlations}
In this appendix we compute the two-time correlation function of the number of photons in a squeezed resonator coupled to a thermal bath and in a coherent state. The Hamiltonian of a squeezed resonator is given by \cref{eq:squeezed_resonator_H} with $g=0$
\begin{align}
\label{eq:squeezed_resonator_only}
H/\hbar=\tilde{\omega}_{r} a^\dag a - \frac{\lambda}{2}\left(a^2 +{a^\dag}^2\right).
\end{align}
For a system described by the  Lindblad master-equation \cref{eq:Lindblad_rho}, the time evolution of an arbitrary operator $O_i(t)$ is given by 
\begin{align}
\begin{split}
&\frac{d}{d t}\left<O_i(t)\right>=\frac{i}{\hbar}\left<\left[H,O_i(t)\right]\right> +\\
&\sum_{m} \left(\left<L_m^\dagger O_i(t) L_m\right>  -\frac12 \left< L^\dagger_m L_mO_i(t) \right> -\frac12 \left< O_i(t)L^\dagger_m L_m \right>\right),
\end{split}
\end{align}	
where $L_m$ are the Lindblad superoperators and $m$ is a positive integer $\lbrace m=1,2,..\rbrace$. If the resonator is coupled to a thermal bath, the Lindblad operators are
\begin{align}
&L_1 = \sqrt{\kappa \left(\bar{n}+1\right) }a\\
&L_2=\sqrt{\kappa \bar{n}} a^\dag       
\end{align}

The time derivative of $\vec{O}(t)$, where $\vec{O}(t) =  \begin{pmatrix} n(t), \ a^{2}(t), \ {a^{\dagger}}^{2}(t), \ \mathds{1} \end{pmatrix}^{T}$, gives rise to a set of coupled differential equations that can be written in the form of \cref{eq:EOM_1op}, with
\begin{align}
G=\left(\begin{matrix}
-\kappa      & -i \lambda         & i \lambda          & \kappa \bar{n}         \\
i2  \lambda  & -i2\tilde{\omega}_{r}-\kappa & 0                  & i\lambda  \\
-i2  \lambda & 0                  & i2\tilde{\omega}_{r}-\kappa & -i\lambda \\
0            & 0                  & 0                  & 0
\end{matrix}\right).
\label{eq:GG}
\end{align}
\subsection{Squeezed resonator} 
For a squeezed resonator coupled to a zero-temperature bath, $\bar{n}=0$, \cref{eq:GG} becomes
\begin{align}
G=	\begin{pmatrix}
	-\kappa      & -i \lambda         & i \lambda          & 0         \\
	i2  \lambda  & -i2\tilde{\omega}_r-\kappa & 0                  & i\lambda  \\
	-i2  \lambda & 0                  & i2\tilde{\omega}_r-\kappa & -i\lambda \\
	0            & 0                  & 0                  & 0
	\end{pmatrix}.
	\label{eq:G0}
\end{align}
To obtain the two-time correlation function $\langle n(\tau)n(0)\rangle$, we use the quantum regression theory, see \cref{ap:QRT}, to obtain the relation
\begin{align}
\label{eq:QRT}
\frac{\partial}{\partial \tau}
\left<\vec{O}(\tau) n(0)\right>
=G\left<\vec{O}(\tau) n(0)\right>.
\end{align}
We solve this equation by diagonalizing $G$ and using the initial conditions for the steady-state solution of $\left<\vec{O}(\tau) n(0)\right>$. To obtain the latter, we compute the equations of motion of the expectation values of the operator 
\begin{align}
\vec{\psi} =\begin{pmatrix} a^\dag a, \
a^2, \
{a^\dag}^2, \
{a^\dag}^2 a^2, \
{a^\dag}^3 a, \
a^\dag a^3, \
a^4, \
{a^\dag}^4, \
\mathds{1} 
\end{pmatrix}^T
\end{align}
and write them as $\frac{d}{d t}
\langle\vec{\psi}(t)\rangle	=
M
\langle\vec{\psi}(t)\rangle	$, where
\begin{widetext}
	\begin{align}
M=	\left(
	\begin{matrix}
	-\kappa &	 -i\lambda & 	i\lambda  &		 0 &		 0 &		 0 &		 0 &		 0 & 0 \\
	2i\lambda  & -2i\tilde{\omega}_r-\kappa &0 	& 0 &		 0 & 	0 & 	0 &		 0 & i\lambda \\
	-2i\lambda &		 0&  2i\tilde{\omega}_r-\kappa  & 0 & 0 & 0 & 0 & 0 & -i\lambda \\
	0 	& -i\lambda 	& 	i\lambda &	-2\kappa &	2i\lambda & -2i\lambda  &		 0 &	 0 &	 0\\
	-3i\lambda 	& 0 	& 		0 &	-3i\lambda &2i\tilde{\omega}_r-2\kappa &  	0 &		 0 &	i\lambda &	 0\\
	3i\lambda 	& 0 	& 		0 &	3i\lambda &	 0 &  -2i\tilde{\omega}_r-2\kappa  &		-i\lambda &	 0 &	 0\\
	0 	& 6i\lambda 	& 		0 &		0 &	 0 &  	4i\lambda &		-2\kappa-4i\tilde{\omega}_r &	 0 &	 0\\
	0 	& 0 	& 	-6i\lambda &		0 &	-4i\lambda &  	0 &		 0 &	-2\kappa+4i\tilde{\omega}_r &	 0\\
	0 	& 0 	& 		0 &		0 &	 0 &  	0 &		 0 &	 0 &	 0
	\end{matrix}
	\right)
\end{align}
\end{widetext}
The steady-state solution is given by the null eigenvector of $M$, defined by $M\langle \vec{\psi}\rangle _\infty=0$, and corresponds to
\cleardoublepage
\begin{align}
\label{eq:ss_squeeze_sol}
\langle \vec{\psi}\rangle
_{\infty}
=
\begin{pmatrix}
\frac{2 \lambda ^2}{4 \tilde{\omega}_r ^2+\kappa ^2-4 \lambda ^2}\\
\\ 
-\frac{\lambda  (2 \tilde{\omega}_r +i \kappa )}{-4 \tilde{\omega}_r ^2-\kappa ^2+4 \lambda ^2}\\
\\ 
-\frac{-2 \tilde{\omega}_r  \lambda +i \kappa  \lambda }{4 \tilde{\omega}_r ^2+\kappa ^2-4 \lambda ^2}\\
\\ 
-\frac{-4 \tilde{\omega}_r ^2 \lambda ^2-\kappa ^2 \lambda ^2-8 \lambda ^4}{\left(4 \tilde{\omega}_r ^2+\kappa ^2-4 \lambda ^2\right)^2}\\
\\ 
\frac{6 \left(2 \tilde{\omega}_r  \lambda ^3-i \kappa  \lambda ^3\right)}{\left(-4 \tilde{\omega}_r ^2-\kappa ^2+4 \lambda ^2\right)^2}\\
\\ 
\frac{6 \left(2 \tilde{\omega}_r  \lambda ^3+i \kappa  \lambda ^3\right)}{\left(4 \tilde{\omega}_r ^2+\kappa ^2-4 \lambda ^2\right)^2}\\
\\ 
\frac{3 \left(4 \tilde{\omega}_r ^2 \lambda ^2+4 i \tilde{\omega}_r  \kappa  \lambda ^2-\kappa ^2 \lambda ^2\right)}{\left(4 \tilde{\omega}_r ^2+\kappa ^2-4 \lambda ^2\right)^2}\\
\\ 
\frac{3 \left(4 \tilde{\omega}_r ^2 \lambda ^2-4 i \tilde{\omega}_r  \kappa  \lambda ^2-\kappa ^2 \lambda ^2\right)}{\left(4 \tilde{\omega}_r ^2+\kappa ^2-4 \lambda ^2\right)^2}\\
\\ 
1
\end{pmatrix},
\end{align}

\begin{figure}
	\centering
	\includegraphics[width=1\linewidth]{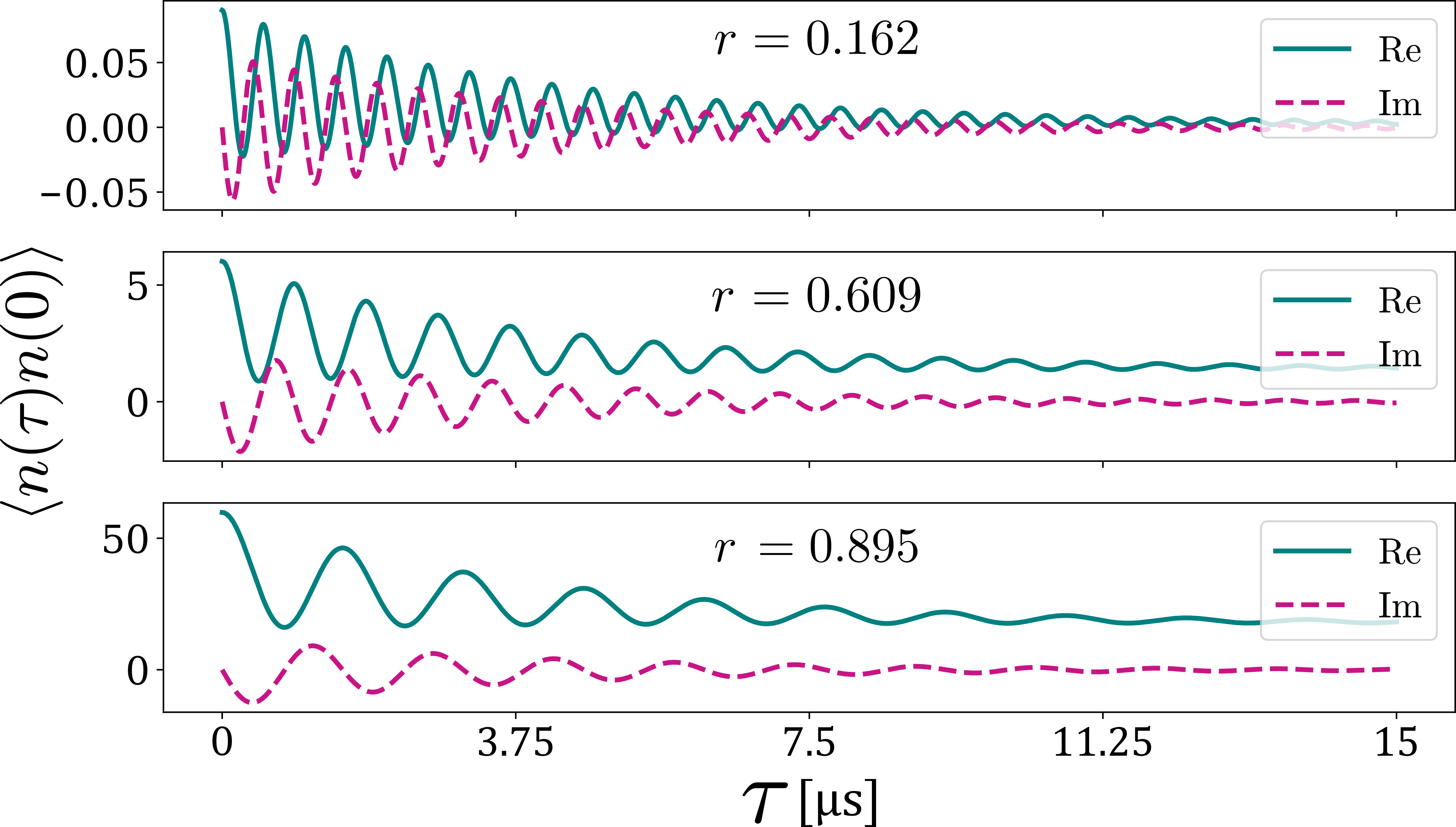}
	\caption{The two-time correlation for the photon-number operator vs $\tau$, in a squeezed state for different values of r (solid blue and dashed purple). This correlation provide an indication for the effective temperature of the system, where the fluctuations increase for larger squeezing parameter $r$.}
	\label{fig:ntaun0}
\end{figure}

From this expression, and using the canonical commutation relations, we obtain the initial conditions of \cref{eq:QRT} for $\tau=0$, namely
\begin{align}
\langle \vec{O}(0)n(0)\rangle_\infty 
=\begin{pmatrix}
\frac{3 \lambda ^2 \left(4 \tilde{\omega}_r ^2+\kappa ^2\right)}{\left(4 \tilde{\omega}_r ^2+\kappa ^2-4 \lambda ^2\right)^2}\\
\\
\frac{2 \lambda  (2 \tilde{\omega}_r +i \kappa ) \left(4 \tilde{\omega}_r ^2+\kappa ^2-\lambda ^2\right)}{\left(4 \tilde{\omega}_r ^2+\kappa ^2-4 \lambda ^2\right)^2}\\
\\
\frac{6 \lambda ^3 (2 \tilde{\omega}_r -i \kappa )}{\left(4 \tilde{\omega}_r ^2+\kappa ^2-4 \lambda ^2\right)^2}\\
\\
\frac{2 \lambda ^2}{4 \tilde{\omega}_r ^2+\kappa ^2-4 \lambda ^2}
\end{pmatrix}.
 \end{align}
Next, we solve \cref{eq:QRT} by diagonalizing $M$ and derive an analytic expression for $\langle n(t)n(0)\rangle$, see Fig.~\ref{fig:ntaun0}. Using this expression, we can compute the integrated two-time correlation defined in Eq.~(\ref{eq:eta}) and used in Fig~\ref{fig:frames_comaprison}, 
\begin{align}
\eta & =\frac{2\lambda ^2 \left(4 \tilde{\omega}_r ^2+\kappa ^2\right) \left(4 \tilde{\omega}_r ^2+5 \kappa ^2-4 \lambda ^2\right)}{\kappa  \left(4 \tilde{\omega}_r ^2+\kappa ^2-4 \lambda ^2\right)^3}.
\label{eq:eta}
\end{align}
To evaluate the dephasing rate of a squeezed resonator, we need to compute the two-time correlation of $\tilde{n}$, see \cref{eq:NplusHalf,eq:gamma_photon}. In order to find this quantity, we compute all permutations obtained from the multiplication between $\tilde{n}(\tau)$ and $\tilde{n}((0))$, i.e. all the matrix elements of the $9\times9$ matrix $\langle \vec{O}^T(\tau) \vec{O}(0)\rangle$. We refer the reader to the Wolfram Mathematica code in the online repository for all the details \footnote{\url{https://github.com/InbarShani2610/Coherence_Properties_of_a_Spin_in_a_Squeezed_Resonator/tree/Mathematica_code}}. This calculation enables us to compute the integrated two-time correlation function $\tilde \eta$ and infer the decoherence rate of a spin coupled to a squeezed resonator, \cref{eq:squeezed_gamma_photon}.
\subsection{Thermal bath}
We now consider a thermal bath in the absence of squeezing ($\lambda=0$). In this case, it is sufficient to compute the expectation value of $\vec{O} = \begin{pmatrix} n(t), \ \mathds{1} \end{pmatrix}^{T}$, whose time derivative takes the form of \cref{eq:EOM_mat} with
\begin{align}
G= \left(\begin{matrix}
- \kappa & \kappa\bar{n}\\
0 & 0
\end{matrix}\right).
\end{align}
The steady-state expectation values are obtained by the null eigenstate of $G$, defined by $G\langle \vec{O}\rangle =0$, leading to $\langle n(t)\rangle_\infty = \bar{n}$. To obtain the two-time correlation  $\langle n(t+\tau)n(t) \rangle _\infty$,  we use the quantum regression theorem \cref{eq:twotime_corr} and obtain
\begin{align}
\frac{d}{d\tau} \left< \begin{matrix} n(t+\tau)n(t)\\ \mathds{1}\; n(t) \end{matrix} \right> = \left(\begin{matrix}
- \kappa & \kappa\bar{n}\\
0 & 0
\end{matrix}\right)\left<\begin{matrix} n(t+\tau)n(t)\\ \mathds{1}\; n(t) \end{matrix} \right>.
\end{align}
Next, we compute the initial conditions (as done previously), and find $\langle n(0)n(0)\rangle=\bar n(2\bar{n}+1)$. Finally, by diagonalizing $G$, we obtain
\begin{align}
\label{eq:thermal_n_corr}
\langle  n(\tau)n(0) \rangle_\infty = \bar{n}^2+\left(\bar{n}^2+\bar{n}\right) e^{- \kappa\left| \tau\right| },
\end{align}
Here, one can check that at $\tau\rightarrow\infty$ the operator  $n(\tau)$ is not correlated to the operator $n(0)$ and $\lim_{\tau \to \infty}  \langle  n(\tau)n(0) \rangle_\infty = \langle  n(\tau) \rangle_\infty \langle n(0) \rangle_\infty  = \bar{n}^2$. Using \cref{eq:thermal_n_corr}, we can compute 
\begin{align}
\label{eq:thermal_delta_eta}
\eta=\int_{-\infty}^{\infty} \langle n(\tau)-\bar{n} \rangle \langle  n(0)-\bar{n} \rangle d\tau = \frac{2\left(\bar{n}^2+\bar{n}\right)}{\kappa}.
\end{align}

leading to the known result \cite{bertet2005dephasing}
\begin{align}
\Gamma_\phi^{\rm photon}=  \frac{4 \chi_0^2\left(\bar{n}^2+\bar{n}\right)}{\kappa}.
\end{align}

\subsection{Coherent state}
To consider the dephasing of a spin coupled to a resonator in a coherent state, we consider the following Hamiltonian and Lindblad operator
\begin{align}
H&=\tilde{\omega}_r a ^\dag a\\
L&=\sqrt{\kappa}\left( a-\alpha e^{i\tilde{ \omega}_r t}\right).
\end{align}
The steady state of this system corresponds to a pure state $\rho(t)=|\psi(t)\rangle \langle \psi(t)|$, where the coherent state $|\psi(t)\rangle = |\alpha e^{i\tilde\omega_r t}\rangle$ satisfies $L|\psi(t)\rangle=0$. Applying the transformation $U=e^{-i\tilde{ \omega}_{r} a^\dag a t }$ on the Hamiltonian and the Lindblad operator above we obtain 
\begin{align}
	\tilde{H}&=0\\
	\tilde{ L} &= \sqrt{\kappa}\left(a-\alpha\right)e^{i\tilde{ \omega}_rt}.
\end{align}
In the new frame, we can write the coupled equations of motion for $\vec{O}$, where $\vec{O} =  \begin{pmatrix} n(t), \ a(t), \ {a^{\dagger}}(t), \ \mathds{1} \end{pmatrix}^{T}$, in the form of \cref{eq:EOM_1op} with 
\begin{align}
G= \left(
\begin{matrix}
-\kappa  & \frac12 \alpha ^* \kappa  & \frac12 \alpha  \kappa  & 0 \\
0 & -\frac12 \kappa  & 0 & \frac12 \alpha  \kappa  \\
0 & 0 & -\frac12 \kappa  & \frac12\alpha ^* \kappa   \\
0 & 0 & 0 & 0 \\
\end{matrix}
\right).
\end{align}
The steady state solutions result in
\begin{align}
\langle\vec{O}  \rangle_\infty=\left(
\left|\alpha \right|^2,~\alpha,~\alpha ^*,~1\right)^T.
\end{align}
In order to find $\langle  n(\tau)n(0) \rangle$ we diagonalize the system and plug in the initial conditions in the steady-state solution for $\tau=0$ (as done previously for \cref{eq:ss_squeeze_sol}). Thus, we obtain
\begin{align}
\langle  n(\tau)n(0) \rangle_\infty = \left|\alpha\right| ^4+\left|\alpha\right| ^2 e^{-\frac12\kappa  \left|\tau\right| }.
\end{align}
We can, again, check that at $\tau\rightarrow\infty$, the operator  $n(\tau)$ does not correlate with the operator $n(0)$  such that $\lim_{\tau \to \infty}  \langle  n(\tau)n(0) \rangle_\infty = \langle  n(\tau) \rangle_\infty \langle n(0) \rangle_\infty  = \left|\alpha\right|^4$. In this case, we obtain
\begin{align}
\eta=\int_{-\infty}^{\infty} \langle n(\tau)-\bar{n} \rangle \langle  n(0)-\bar{n} \rangle d\tau =\frac{  4\left|\alpha\right|^2}{\kappa}.
\end{align}
By changing the notation $\left|\alpha\right|^2$ to $\bar{n}$, we obtain the dephasing rate due to photon noise in a coherent-state \cite{blais2004cavity}
\begin{align}
\Gamma_{\phi}^{\rm photon} = \frac{8\chi^2\bar{n}}{\kappa}.
\end{align}

\bibliography{Ref}
\end{document}